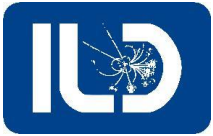

# Beyond the CMOS sensors: the DoTPiX pixel concept and technology for the International Linear Collider[A]

Nicolas T. Fourches*,
*CEA/Saclay and University Paris-Saclay,
Geraldine Hallais**, Charles Renard**
**CNRS/C2N University Paris-Saclay

## Abstract

CMOS sensors were successfully implemented in the STAR tracker [1]. LHC experiments have shown that efficient b tagging, reconstruction of displaced vertices and identification of disappearing tracks are necessary. An improved vertex detector is justified for the ILC. To achieve a point (spatial single layer) resolution below the one-µm range while improving other characteristics (radiation tolerance and eventually time resolution) we will need the use of 1-micron pitch pixels. Therefore, we propose a single MOS transistor that acts as an amplifying device and a detector with a buried charge-collecting gate. Device simulations both classical and quantum, have led to the proposed DoTPiX structure. With the evolution of silicon processes, well below 100 nm line feature, this pixel should be feasible. We will present this pixel detector and the present status of its development in both our institution (IRFU) and in other collaborating labs (CNRS/C2N).

**Introduction:**
The future International Linear Collider Experiments require some progress in the inner detector when compared to hadron experiments, especially in spatial resolution (For review of the concerning the impact on physics see [1] [2]). Quite some of R&D was done under the auspices of the MIMOSA CMOS sensors developments. On prototype pixel arrays the resolution reached the 5-micrometer point resolution [3] [4]value or more recently below [5]. The material budget will be reduced by the use of thinned silicon substrates that can be bent if the radius is not too low. There are in fact two separate needs. The first is spatial resolution and the second is time resolution. Here we state that close detector layers, one for timing, and the other for particle-hit evaluation can handle the two needs separately. This can be possible with the ILC constraints that do not impose a trigger and therefore the pixel detector can be read continuously during the bunch train. The constraints on the speed of the readout can be alleviated and we can focus on the granularity with the aim of a pixel pitch lower than one micrometer. On the physics side, the proposed pixels will increase the ability of the vertex detector to resolve close tracks with an improved evaluation of the vertices. Moreover, multiple hits within a single pixel can be reduced in number. Separation of tracks can then be made with a resolution close to the pitch between pixels.

**Physics**
The probe of physics at short lengths is currently done by increasing the energy. Precise measurements for b-tagging should be necessary for the detection of dark matter $\chi$ particles [6] where $e+e- \rightarrow \bar{\chi}\chi\gamma$ is not the only channel to explore, for a study see reference [7].

$$e+e- \rightarrow \bar{\chi}\chi + Z (\rightarrow jj, \mu^+\mu^-, e^+e^-) \qquad (1)$$

where *j* refers to a jet arising from any of *u, d, s, c, b* quarks
In addition, the Dark Sector could be probed by the indirect detection of the possible mediator particle, in the Higgs sector [7] or in the case of monophoton processes [8] [9], by assessing the absence of charged particles close to the interaction point.

$$e+e- \rightarrow \bar{\chi}\chi\gamma \qquad (2)$$

In order to get some information the Z should decay as stated in equation (1) in visible particles (hadron shower) that should benefit from the tagging of light hadrons quarks. (Heavy) [10] Flavor physics could profit from the proposed vertex detector by discriminating secondary vertices from the interaction point. For instance, the decay distance that could be directly probed by such a detector is of one micrometer. However, the effect of statistics would allow probing lower decay lengths by the effect on the lifetime distribution. Direct detection of short-lived particles will be facilitated. The probe of invisible decays (the so-called dark sector [9] will be facilitated by the implementation of such vertex detectors.
As shown by established studies on the ATLAS experiment Higgs Physics and Top Physics benefit from an efficient b-tagging of jets with high transverse momentum [11]. In this case, it allows the direct determination of the lifetime of some hadron with high lifetime; in the case of the proposed vertex detectors, we could go to lower lifetime evaluation, which means that we could improve short-lived hadron tagging. Additionally c-hadrons and $\tau$ leptons tagging is one of the objectives of the ILC. A study of vertexing methods for the ILC was made in [12].

**Detector:**
The operation of the pixel vertex detector at ILC strongly depends on the machine parameters. One should recall the beam parameters of the proposed ILC. The frequency is 5 Hz (200 ms period), with a pulse containing 2625 bunches separated by 369.2 ns, this gives a 0.969 ms beam pulse duration and a bunch length=300 μm. We have chosen to make a table similar to that of the Letter Of Intent [13] (Table 1).

| Layer | Radius in mm | Number of Bunchs Crossings and corresponding duration | Pixel Occupancy In brackets (25 x 25 μm x μm pixels) |
|---|---|---|---|
| 0 | 16 | 83 (30 μs) | 0.53 % (3.33 %) |
| 1 | 17.9 | 83 (30 μs) | 0.3 % (1.90 %) |
| 2 | 37 | 333 (123 μs) | 0.06% (0.40 %) |
| 3 | 38.9 | 333 (123 μs) | 0.053% (0.33 %) |
| 0 | // | 2625 ( 1ms) | 16.7% |
| 1 | // | 2625 | 9.48% |
| 2 | // | 2625 | 0.47% |
| 3 | // | 2625 | 0.42% |

Table 1: Updated from the letter of intent, International Large Detector, background improvements due to the reduction of the area of the pixel by a factor 625 [13] (page 32 ILD LOI). This occupancy value show that these pixels alleviate the need to have time stamping ability. We recall the time structure of the pixel. The LOI (Letter Of Intent) figures were based on a 25 μmx 25 μm sized pixel (MIMOSA 8 like) [14].

This sets the operating conditions of the proposed small pixel option, for the first two layers read out just after the train induces an occupancy, which is relatively high. For the outer layers the occupancy is sufficiently low. However, the original figures assume a 10-pixel cluster size, which will be significantly reduced to a single pixel, with the small pixels where charge spreading should be negligible (see V. Kumar et al. M2 report, 2016, IRFU, on request, for GEANT4 simulations and [15] [16]). Then the occupancy could be reduced to a few percent. The charge spreading in CMOS sensors is mainly due to the diffusion of the carriers with a much reduced drift component in the carrier transport process. To mitigate these thin and depleted sensors should be considered as an alternative. We have introduced the DoTPiX architecture to meet these requirements. This pixel type is sufficiently thin to withstand full depletion, and simulation results have resulted in drift-dominated transport based device.

**Track reconstruction and vertex determination: simple considerations**
Most of the problems in track reconstruction have their origin in multiple scattering [17]. However, progress has been done to mitigate this problem in silicon trackers. We give one example where improvements can be seen. We give prelaminary estimations: let us take simple considerations on the inner layers in the binary pixel mode.

- e is the thickness of the layer and L the distance between layers for a particle with θ polar angle equal to π/2 , $L_{12}$ is the distance between the first and second layer, $L_{01}$ is the distance between the IP and the first layer.
- The maximum deflection angle $α_{max}$ is given by the error on the impact δr of the particle on the layer δr= $α_{max}$ x e , hence $α_{max}$ = δr/e .In our case δr < 1 µm , from GEANT4 (V.Kumar) simulations. Therefore, we can resolve impacts better than one micron. We have: δr < δd.
- With the first two layers, we can estimate the maximum (worst case) error IP determination.
- On the first layer the maximum error on the impact position is $δd_1$ with $δd_1$ =$α_{max1}$ x $L_{01}$
- The error on the impact on the second layer is within $δd_2$= $α_{max2}$ x $L_{12}$ , If we can resolve $δd_2$ on the second layer. There is also a similar $α_{max}$ deflection angle after the particle traverses the second layer. In any case: $δd_2$ ≈ $δd_1$ .
- We have ($δd_2$ + $δd_1$)/ $L_{12}$ ≈ 2x δd/ $L_{12}$ =$α_{max\,w}$, worst case deflection angle.
- ΔIP is the error on the estimation of the interaction point position and is given by:
- ΔIP ≈ $L_{01}$ x $α_{max\,w}$ ≈ 2 x $L_{01}$ x δd/ $L_{12}$ worst case and because $L_{12}$ ≈ $L_{01}$, this leads to ΔIP ≈ 2 x δd this is a limit only dependent on the point resolution of the layer. With δd = 1 µm point resolution , with a δd=1 µm a this gives a worst case ΔIP < 3 µm with a two pixel layers reconstruction.
- This means that the reconstruction with several more layers put in the vertex detector the determination of the IP and the primary and secondary vertex could reach an a precision, which has not been achieved up to now.

Of course, we normally have to take into consideration the curvature of the tracks due to the magnetic field, but with extra layers this should improve resolution in vertex and IP determination compared with this simple 2-layer approach. In a recent study, we have made a more thorough analysis of the resolution using some simulation and reconstruction. These preliminary studies based on mean square linear regression show that reducing the pitch of the pixel has a very positive effect on the track reconstruction [15]. Recently, vertexing packages have been introduced [18].

**Motivations for a reduced pitch pixel:**
The optimal resolution of a pixel of pitch p is given approximately by the relation: σ = p x 0.15 in unit of length so [19] the 25 µm large MIMOSA pixels could reach as we had shown 2.5x1.5 = 3.75 µm (we measured an estimated value of : 5 µm) with a clustering scheme (see ref for a summary [20]). With no clustering scheme and with a binary readout [21] the resolution for a square is $p/\sqrt{12}$ ~ p x 0.29 which means that a point resolution of 0.29 µm or less may be reached. We gain a factor of 10 with a much simpler readout with no charge sharing reconstruction scheme. It was stated in [19] that the δ rays (secondary electrons) were the limiting factor rendering submicron resolution unreachable. As the proposed pixel is designed to eliminate charge sharing and more importantly as it is based on a thin detecting layer and planned to be implemented in a less than 50 µm substrate, this effect is expected to be reduced. Preliminary GEANT4 simulations were made in Table 2 [16]. In the advent of neighboring pixels nonetheless being hit, a reconstruction algorithm could be probably be developed to eliminate this problem.
The first integrated CMOS pixel sensors have been implemented in the STAR experiment [22] In terms of spatial resolution, the goal of sub 5 microns has been reached in practice. Increasing the density of pixels has reduced effect on the data flow as shown in [16] . When zero suppression is implemented the data flow only depends of the log of the density of pixels and hence is sublinear. Zero suppression can be easily implemented with standard CMOS processes.

**Principle of the DoTpIX:**

The DoTpiX (from Dot (quantum Dot pixel) in its proposed form is a single MOS n-channel transistor supplemented by a buried quantum box that may evolved to a quantum dot in its final form. The role of the buried channel is two-fold. First, it acts as a carrier (holes in this case) collection device and second as a second channel control gate. Simply expressed this makes a very compact pixel based on a single MOS structure with a minimum control a readout outputs and inputs (Figure 1).
In the proposed device, the buried gate is a quantum well for holes and a small barrier for electrons. It is made of an epitaxial Ge layer with a thickness of 20 nm; this layer is normally strained compressively. The operation of this proposed pixel device [23] was described in [24] and several simulations have assessed the validity of the principle. In addition, the device feasibility in terms of material engineering is currently being established both by simulations with commercial codes like ATHENA$^©$ (Silvaco) and by experimental tests of growth processes and oxidization among others aspects. Thermal budget and Ge and Si exodiffusion are critical for the process, if we go to a fully CMOS compatible device.

**Sequential operation of the n-channel DoTPiX**
 Simulations with commercial packages like Silvaco ATLAS in quantum mode have been extensively used for the evaluation of the device. These suggest that a readout time down to 20 ns may be reached for each pixel individually. A full readout with a pixel-to-pixel scheme will take more than 200 microseconds for 100 x 100 pixel =10000 pixels arrays. This suggest that a row scan with parallel bottom of column readout is preferable. This kind of readout with zero suppression was introduced in other CMOS arrays and are perfectly compatible with the DoTPiX if implemented in a CMOS process. In this case, the total readout time will be reduced to two microseconds. To summarize the main sequential operation.
- In the detection mode the gate of (all) pixels are set to –Vg (negative voltage) with respect to the substrate, with a similar bias for the drain. The source is set to a negative value as well.
- In readout mode the gate is set to a positive voltage (this can be +Vdd=3.3 V) and the source is connected to a current bias.

One should note that all the drains of a single column are connected, and all the sources are as well. In readout mode only a current bias is set for all the sources, a single transistor is connected by switching it on using a positive upper gate bias.
Injecting carriers in the Ge layer is the way to reset the device. However, due the natural resetting time of this pixel design, this should not be necessary.

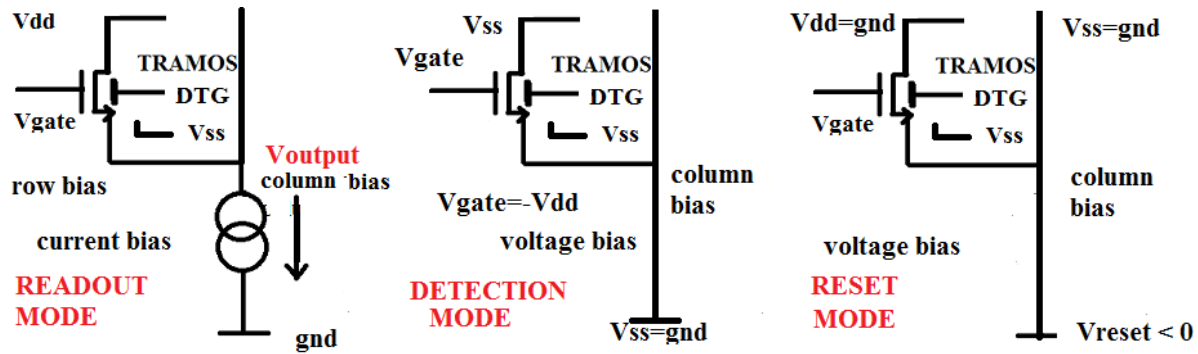

Figure 1: single DoTPiX pixel and connections, with the electric operation mode [24]

We use a column vertical line for the sources and so for the drains (alternatively, a row line can be used for the drains) a row line connects the upper control gates. Then the operation of each device is made easy. The proposed device has been simulated in an early study. The simulation results are summarized in Table 2.

| L (µm) Gate Length | Buried Gate transconductance (µA/V) | Buried Gate-Upper Gate capacitance | ΔV (buried gate) for 80 e charge deposited | ΔV (source) for 80 e charge deposited and a 5 kohms source resistance | Conversion Factor in µV/e (source of the transistor) 5 kohms source resistance |
|---|---|---|---|---|---|
| 1 µm | 52 | 5.31 fF | 3.01 mV | 785 µV | 9 µV/e |
| 0.5 µm | 60 | 2.65 fF | 6.00 mV | 1.56 mV | 19.5 µV/e |
| 0.25 µm | 55 | 1.32 fF | 12.0 mV | 3.3 mV | 41.25 µV/e |
| 0.1 µm | 60 | 0.531 fF | 24.1 mV | 7.23 mV | 90.4 µV/e |

Table 2. Summary of early device simulation and computed results for the DoTPiX. A C script allow the generation of e-h pair along a track in the bulk of the device.

In addition, the pixel should have the ability to have a strong memory effect. See Figure 2. The minimum time for readout the signal is approximately 5-10 ns for one single pixel as simulated.

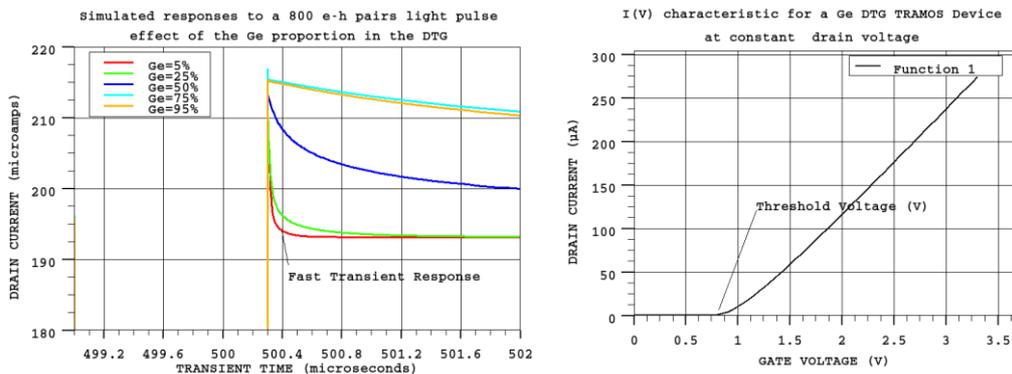

Figure 2: Right normal operation of the n-channel MOS device, drain current versus gate voltage. Left operation of DoTPiX in detection and in further readout mode. The photo-generated e-h corresponds to a value of 800 electron-hole pairs, and the drain current maximum variation 20 µA (from [24]). The percentage is the proportion of Ge in the buried layer.

**Material science challenges and progress:**
If we ignore the buried gate, the DoTPIX reduces to a single n-channel MOS device. This makes it easily feasible using a standard MOS process. We still have to consider how to introduce a buried layer composed of high proportion of germanium. Ion implantation at high energy (up to 1 MeV) has enabled us to evaluate the properties of the buried layers such as the inter-diffusion of Ge and Si species in the neighboring layers. Optimizing the thermal budget strongly reduces this effect. We make an epitaxial silicon layer on the silicon substrate (p-type high resistivity). This is a buffer layer. Then we grow a Ge layer with a thickness of 20 nm. This Ge layer is covered with an epitaxial layer of 20 nm or more of p-type Si. This results in a Si/Ge/Si structure.

We have recently done several tests with UHV-CVD growth technique. We improved surface roughness and the quality of the Ge layer by using low temperature (330 °C°) epitaxial process. Initial cleaning of the silicon surface is also key to improvement. See Figure 3.

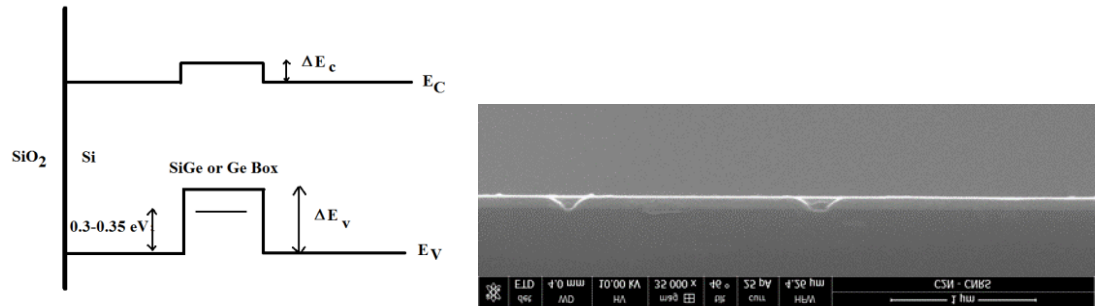

Figure 3: left: Quantum well band diagram for the Ge in Si structure. See reference [25] for details. Right: Ultra-high vacuum CVD growth Si Buffer SiH4 flux=7sccm, P = 3X10$^{-4}$ Torr   Temperature 700°C Ge deposition GeH4 flux =10sccmP = 3.10$^{-3}$ Torr, Temperature: 330°C. This is a SEM view (scanning electron microscope)

**Conclusions:**

At this stage of the R&D, the characteristics of the proposed pixel detector cannot be fully estimated. We are planning to make a test vehicle, which would be controlled by and external circuit. An array of approximately 100 x 100 pixels multiplied by 4 will be used. In this case to avoid complication in readout methods, a column by column and row by row scan will enable the sequential readout of the arrays of pixels. It will be necessary to design a testing bench and an array. This array will be read out in analog mode for the preliminary laboratory tests. We expect starting these tests in 2022. Some important work remains in the control of the process, especially to make the Ge epitaxial substrates compatible with MOS process. This includes many issues such as the thermal budget and mask alignments (auto-aligned MOS transistors are necessary) and the number of interconnecting metal layers available. When all this is overcome, extensive tests can be envisaged. Additionally we will study alternative oxides to improve the radiation hardness.

Note: Up to now, the readout of each single pixel should be made at 10 microseconds or less time interval. This corresponds to 27 bunches. The readout technique, which is possible for CMOS MAPS, would be a repeated scan of the pixel arrays, with a time reference at the start of each scan allowing this. We should know that this technique would blind a changing single row and a single column permanently. Fast readout may be used with all pixels blinded when possible. Alternatively, one could introduce a memory cell (a single capacitor) at the expense of size reduction on the DoTpIX to allow successive memory scans and allow a readout after the bunch train.